\documentclass[fleqn,usenatbib]{mnras}

\usepackage[T1]{fontenc}

\DeclareRobustCommand{\VAN}[3]{#2}
\let\VANthebibliography\thebibliography
\def\thebibliography{\DeclareRobustCommand{\VAN}[3]{##3}\VANthebibliography}

\usepackage{graphicx}	
\usepackage{amsmath}	

\title[Deuterated water abundance in RCW\,120~S2]{Deuterated water abundance in the young hot core RCW\,120~S2}

\author[M. S. Kirsanova \& A. A. Farafontova]{
Maria. S. Kirsanova,$^{1,2}$\thanks{E-mail: kirsanova@inasan.ru}
Anastasiia A. Farafontova$^{1}$\\
$^{1}$Institute of Astronomy, Russian Academy of Sciences, ul. Pyatnitskaya 48, Moscow, 119017, Russia\\
$^{2}$Astro Space Centre, Lebedev Physical Institute, Russian Academy of Sciences, ul. Profsoyuznaya 84/32, Moscow, 117997, Russia\\
}

\date{Accepted June 02, 2025}

\pubyear{2025}
\begin{document}
\label{firstpage}
\pagerange{\pageref{firstpage}--\pageref{lastpage}}
\maketitle

\begin{abstract}
Emission of water molecules cannot be observed from Earth, less abundant isotopologues, such as H$_2^{18}$O and HDO, are used to trace water in star-forming regions. The main aim of this study is to determine HDO abundance in the hot core RCW\,120~S2. We performed observations of the hot core in the 200-255~GHz range using the nFLASH230 receiver on the APEX telescope. Two HDO lines were detected toward RCW\,120~S2. Their intensities are described by excitation temperature $\approx 290$~K and gas number density $\geq 10^9$~cm$^{-3}$. The emission originates from the hot core rather than the warm dense envelope surrounding a central young stellar object. The HDO column density ranges from $(3.9-7.9)\times10^{13}$~cm$^{-3}$ with the best-fit model value of $5.6\times10^{13}$~cm$^{-3}$. The HDO abundance relative to hydrogen is $1.7\times10^{-9}$. This HDO abundance value is among the lowest reported for hot cores. Combined with the non-detection of the H$_2^{18}$O line, we conclude that protostellar heating in RCW\,120~S2 is still in its early stages.
\end{abstract}




\section{Introduction}\label{sec:intro}

The formation of water in the Universe attracts significant scientific interest, as this molecule is fundamental to all known life. {\it Herchel} space telescope performed extensive observations of water emission lines, enabling researchers to identify the primary formation pathways for both water and its deuterated isotopes. The main results of the WISH program on {\it Herschel} summarised by \citet{vanDishoeck21}. 

Both water and HDO molecules primarily form on the surfaces of dust grains within molecular clouds, prior to the onset of active star formation \citep{2013ChRv..113.9043V}. The surface main reactions for the water and HDO are ${\rm H} + {\rm OH} \rightarrow {\rm H_2O}$ and ${\rm D} + {\rm OH} \rightarrow {\rm HDO}$, respectively, see~\citet{1982A&A...114..245T}. Reactions of ${\rm OH} + {\rm H_2/HD} $ and ${\rm OH/OD} + {\rm H_2}$ can also play a significant role, see \citet{2007ApJ...668..294C, 2012ApJ...749...67O}. The abundance of D atoms {\rm accreting} from the gas on the grain is higher compared with that of H atoms; therefore, the ratio of HDO/H$_2$O grows with time in the cold medium. In the gas phase, reactions of ${\rm H_2D^+}$ and HD with molecular ions containing oxygen lead to enhancing the HDO abundance. Some HDO can also be formed in the gas phase via reactions ${\rm H_2D^+} + {\rm H_2O} \rightarrow {\rm H_3^+} + {\rm HDO}$. In cold molecular clouds, the gas-phase abundance of H$_2$D$^+$ increases relative to H$_3^+$ over time, providing a continuous source of HDO. Because of all these processes, the HDO/H$_2$O ratio becomes higher than the overall [D]/[H] ratio found in the local interstellar medium.

During the hot core/corino phase, when rising dust temperatures lead to mantle evaporation, abundant water and HDO are released into the gas phase, making their emission observable. Statistical analysis of water and HDO abundances in hot cores serves two key purposes: calibration of astrochemical models and constraining the timescales of deuteration processes. Therefore, each new measurement of water and HDO abundances provides critical data that tracks deuteration efficiency across evolutionary stages and advances our understanding of interstellar water chemistry.

Unfortunately, the ground-state and low-excitation water lines (upper state energies of the particular transitions $E_{\rm u} <100$~K) are invisible using ground-based telescopes due to atmospheric absorption. Therefore, less abundant isotopologues, such as H$_2^{18}$O and HDO, are used to trace water in star-forming regions. HDO molecules experience much less absorption and have several lines in the sub- and millimetre range. Early studies of the HDO emission in hot cores can be found in \citet{Jacq_1990}.

The main aim of this study is to determine HDO abundance in the hot core RCW\,120~S2. Young stellar object (YSO) RCW\,120~S2 of $L_{\rm bol}=1163$~$L_\odot$ observed by {\it Herschel} in the mid- and far-infrared (see \citet{2017A&A...600A..93F}, they call it Core~2 but we use RCW\,120~S2 or just S2 hereafter). High-resolution ALMA imaging has resolved the YSO into a multiple system consisting of five distinct point sources as it was shown by \citet{2018A&A...616L..10F}. In their analysis of RCW\,120~S2, \citet{Kirsanova_2021} estimated the temperature of the dense warm envelope and also identified hot core signatures through high-excitation CH$_3$CN lines (upper state energies $E_{\rm u} >100$~K). They also found low abundances of CH$_3$CN and methanol, concluding the hot core is in an early evolutionary stage. \citet{2024paper1} estimated a total hydrogen column density of $N({\rm HI}+2N({\rm H_2}))=3.7\times10^{22}$~cm$^{-2}$ toward the YSO RCW\,120~S2 and detected dozens of rotational lines from CH$_3$OH, CH$_3$CCH and CH$_3$CN.  Their analysis revealed an onion-like emission structure, with CH$_3$CN tracing warmer gas near the YSO ($\approx 60$\,K) compared to the other molecules ($\approx 40$~K). The distance to RCW\,120 is $1.3\pm0.4$~kpc~(Russeil~et~al.,~\cite{2003A&A...397..133R}), therefore RCW\,120~S2 is one of the closest hot cores to the Sun.

\section{Observations and Methods}

The observations were carried out using the APEX telescope~\cite{2006A&A...454L..13G} in Chile on 19-21 of May, 4-5 and 24 June, 22-23 and 30 August 2022, as project E-0109.C-0623A-2022 (PI: Kirsanova M.~S.) within the ESO operated share. The primary objective of this observational project was to acquire high-quality spectra with a noise level of 4~mK at a spectral resolution of 0.3~km\,s$^{-1}$ (Farafontova et al., in prep.). 

The receiver used was the nFLASH230~\cite{2018A&A...611A..98B}. In this article, we present results of three observational setups covering the sidebands: 202.8-206.8, 241-245, and 253-257~GHz. 

The data were calibrated to antenna temperature in real-time using the standard {\it apexOnlineCalibrator} package, but we later additionally applied factors of $\eta_{\rm mb} = 0.81$ and~0.73 for May/June and August, respectively\footnote{https://www.apex-telescope.org/telescope/efficiency/index.php}, to scale the data at the main beam temperature scale. The spatial resolution of the data was 26\arcsec{} at 241~GHz, corresponding to 0.17~pc at the distance of RCW\,120.

We observed position at $\alpha=17\rm ^h$12$\rm ^m$08.700$\rm ^s$ and $\delta=-38^\circ$~30\arcmin47.4\arcsec (J2000). The observations were carried out in position-switching mode with the off-position at $\alpha=17\rm ^h$12$\rm ^m$08.000$\rm ^s$, $\delta=-38^\circ$~36\arcmin03.00\arcsec (J2000). The PWV level ranged between $0.3-0.9$~mm, $T_{\rm sys} \approx 100$~K. For the data presented in this article, total observational time  was 3.5$^{\rm h}$ for each of the frequency ranges.

The baseline correction was performed using the {\tt arpls} function from the {\tt pybaselines} package \citep{erb_2024}. We identified molecular lines using the data from the JPL database~\citep{1998JQSRT..60..883P} for the HDO lines~\citep{1984JMoSp.105..139M} and using CDMS~\citep{2001A&A...370L..49M} for the H$_2^{18}$O line~\cite{1972PhRvA...6.1324D}. For the data analysis, we used both LTE and non-LTE methods. For the LTE analysis, we employed the rotational diagram method, the general formulation of which is described by \citet{Goldsmith1999, 2016ARep...60..702K}. The non-LTE analysis was performed using RADEX software~\citep{vanderTak_2007} with collisional coefficients by \citet{2012MNRAS.420..699F}.

\section{Results}\label{sec:results}

\begin{figure*}
    \centering
    \includegraphics[width=0.7\linewidth]{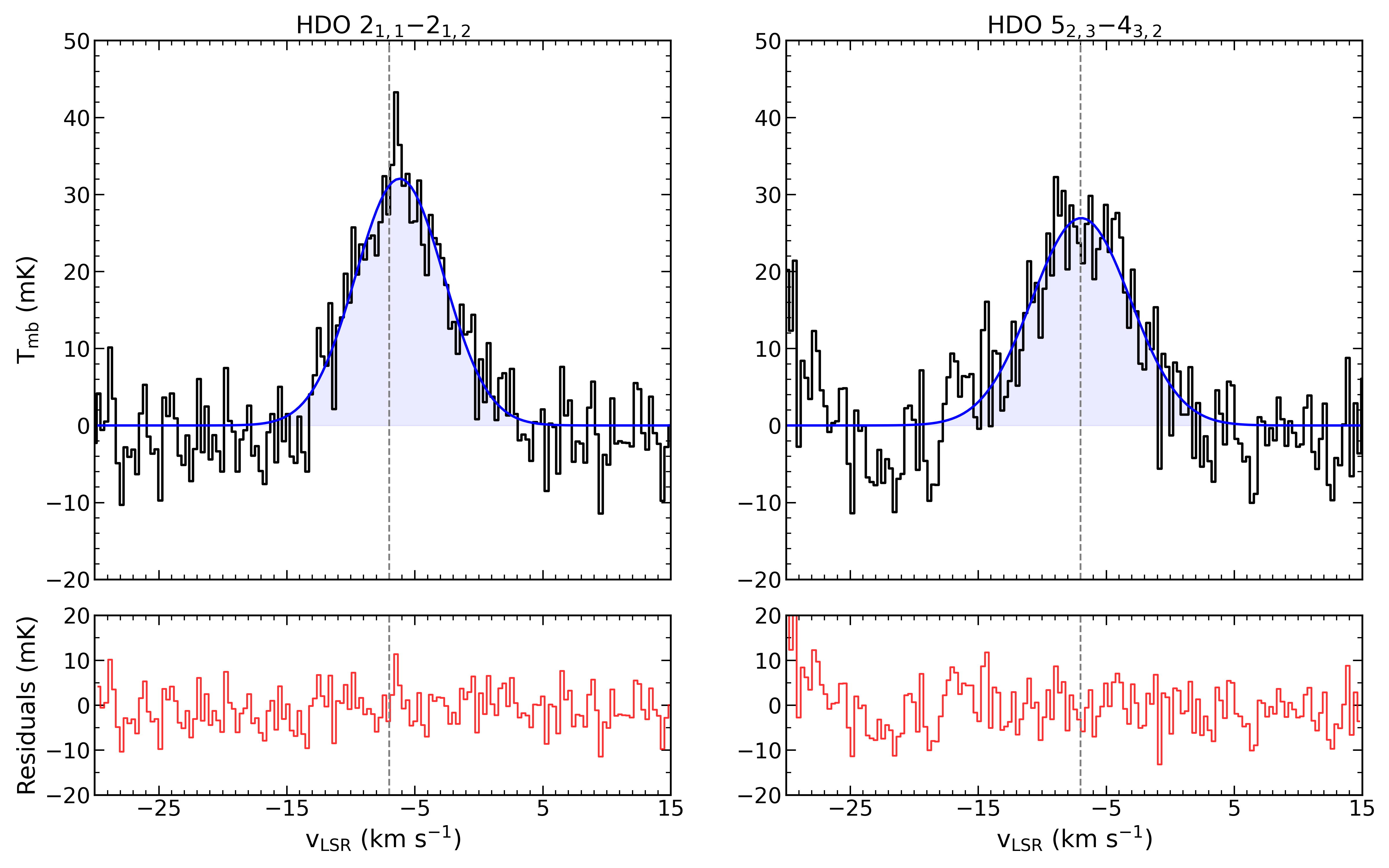}
    \includegraphics[width=0.7\linewidth]{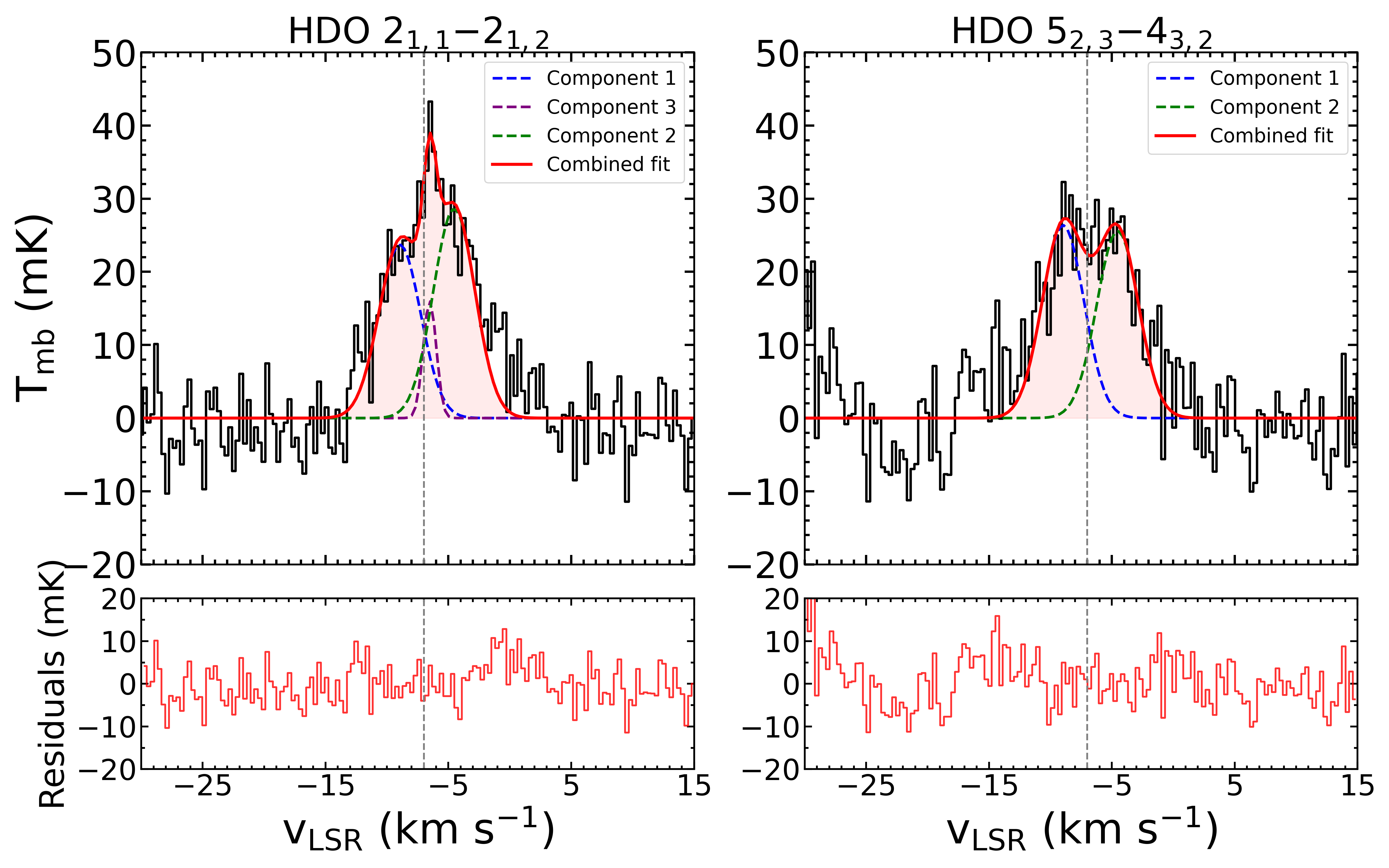}
    \caption{HDO lines toward RCW\,120~YSO\,S2 (black) and rms levels (red). Vertical line shows $V_{\rm LSR}=-7$~km\,s$^{-1}$. Two top panels show the fit by a single Gauss function and the residual spectrum after the fitting procedure. Two bottom panels show the fit by multiple Gauss functions   and also the residual spectrum. }
    \label{fig:HDO_spectra}
\end{figure*}

We detected two HDO lines: $2_{1,1} - 2_{1,2}$ at 241561.55~MHz and $5_{2,3}-4_{3,2}$ at 255050.26~MHz, see Fig.~\ref{fig:HDO_spectra}. Spectroscopic parameters of the lines are shown in Table~\ref{tab:param}. We fitted both the lines by Gauss function and show the fit in Table~\ref{tab:gauss_first}. The range of the $E_{\rm u}$ for the detected lines is $95-437$~K, indicating the presence of gas heated by the embedded YSO. The widths of the $2_{\rm 1, 1}-2_{\rm 1, 2}$ and $5_{\rm 2, 3}-4_{\rm 3, 2}$ {are almost the same. Namely, the width of the $5_{2,3}-4_{3,2}$ is only 12\% higher than the width of the  $2_{1,1} - 2_{1,2}$ line}. The H$_2^{18}$O line at 203407.52~MHz was not detected with the noise level of $\sigma \approx 4$~K. We use $2\sigma$ value as a low limit for the brightness of this line.

\begin{table}
\centering
\begin{tabular}{cccccc}
\hline
&$f$       & $J_{K_a,K_c}$             & $E_{\rm u}$ & $A_{ul}$    & $g_{\rm u}$ \\
&(MHz)     &                             & (K)         & (cm$^{-1}$) & \\
\hline 
H$_2^{18}$O & 203407.52 & $3_{\rm 1, 3}-2_{\rm 2, 0}$ &  203.7 & -5.318   & 7 \\
HDO & 241561.55 & $2_{\rm 1, 1}-2_{\rm 1, 2}$ & 95.2        & -4.926      & 5 \\
HDO & 255050.26 & $5_{\rm 2, 3}-4_{\rm 3, 2}$ & 437.4       & -4.748      & 11\\
\hline
    \end{tabular}
    \caption{Specrtroscopic parameters of the observed lines.}
    \label{tab:param}
\end{table}

\begin{table*}
\centering
\begin{tabular}{cccccc}
\hline 
&$f$             & $T_{\rm mb}$ & $\Delta$v      & $\int T_{\rm mb}$dv & $V_{\rm LSR}$   \\
&   (MHz)               & (mK)          & (km\,s$^{-1}$) & (mK\,km\,s$^{-1}$)   & (km\,s$^{-1}$)  \\
\hline
H$_2^{18}$O & 203407.52 & $\leq 8$ & & & \\
HDO & 241561.55 & $32 \pm 1$   & $8.1 \pm 0.4$ & $280 \pm 10$ & $-6.2 \pm 0.2$  \\
HDO & 255050.26 & $27 \pm 1$   & $9.1 \pm 0.5$ & $260 \pm 10$ & $-7.0 \pm 0.2$ \\
\hline
    \end{tabular}
    \caption{Parameters of the line fit by a single Gauss function.}
    \label{tab:gauss_first}
\end{table*}

We started the analysis of the HDO line emission using the LTE approach and estimated gas kinetic temperature ($T_{\rm kin}$) and HDO number density ($N_{\rm HDO}$) using the integrated intensities $\int T_{\rm mb} dv$ from the fit above. The LTE analysis gives $T_{\rm kin}=286\pm2$~K and $N_{\rm HDO} = (10.4 \pm 1.5)\times 10^{13}$~cm$^{-2}$. The derived $T_{\rm kin}$ exceeds that of the dense warm envelope surrounding the YSO  (see Sec.~\ref{sec:intro}). We therefore conclude that the HDO line emission originates in the hot core.

The $5_{\rm 2, 3}-4_{\rm 3, 2}$ line displays a double-peak structure. This motivated a two-component Gaussian fit with fixed line widths, whose velocity parameters were subsequently adopted to decompose the $2_{\rm 1, 1}-2_{\rm 1, 2}$ line into two components (labeled as~1 and~2 in Figure~\ref{fig:HDO_spectra}). The $2_{\rm 1, 1}-2_{\rm 1, 2}$ transition exhibits enhanced central emission, which we modeled with an additional Gaussian component (component~3). The derived fitting parameters are presented in  Table~\ref{tab:gauss_sec}. The double-peaked structure, represented by components~1 and 2, likely traces the gas kinematics within the hot core, as we discuss below. We note the $2_{\rm 1, 1}-2_{\rm 1, 2}$ and $5_{\rm 2, 3}-4_{\rm 3, 2}$ lines could have red wings, but their brightnesses are about 2 and 1~$\sigma$ level, respectively. Thus, these wings require more sensitive observations for detection.

\begin{table*}
\centering
\begin{tabular}{cccccc}
\hline 
$f$            &  component & $T_{\rm mb}$   & $\Delta$v      & $\int T_{\rm mb}$dv & $V_{\rm LSR}$   \\
(MHz)                    &           & (mK)            & (km\,s$^{-1}$) & (mK\,km\,s$^{-1}$)   & (km\,s$^{-1}$) \\
\hline
241561.55                 & 1         & $23.7 \pm 1.8$ & $4.0 \pm 0.3$ & $101 \pm 28$    &$-9.0 \pm 0.4$\\
                          & 2         & $28.6 \pm 1.9$ & $4.0 \pm 0.3$   & $122 \pm 24$ &$-4.5 \pm 0.4$\\
                          & 3         & $16.0 \pm 3.6$ & $1.3 \pm 0.6$   & $21 \pm 10$ &$-6.5 \pm 0.1$ \\
\hline
255050.26                 & 1         & $26.4 \pm 5.9$ & $4.0 \pm 0.3$ & $110 \pm 30$ & $-9.0 \pm 0.4$ \\
                          & 2         & $25.6 \pm 5.9$ & $4.0 \pm 0.3$& $110 \pm 30$  &$-4.5 \pm 0.4$\\
\hline
\end{tabular}
\caption{Parameters of the line fit by multiple Gauss functions: three for the line at 241561.55 and two for the line at 255050.26~GHz.}
\label{tab:gauss_sec}
\end{table*}

We employed non-LTE radiative transfer analysis to determine the core's physical properties, using the velocity-integrated intensities of the $2_{\rm 1, 1}-2_{\rm 1, 2}$ and $5_{\rm 2, 3}-4_{\rm 3, 2}$ lines extracted from the individual profiles of component~1 and component~2, see Table~\ref{tab:gauss_sec}. The best-fit model and the values of the parameters corresponding to different significance levels with $\chi^2$-test. Implementation of this test to astrophysical problems can be found in \citet{wall_jenkins_2003}.

For our RADEX non-LTE modeling, we adopted 2.7 K for background temperature, while exploring a grid of $n_{\rm H_2}$ ($10^7 -10^{13}$~cm$^{-3}$), $N_{\rm HDO}$ ($10^{11}-10^{16}$~cm$^{-2}$) and $T_{\rm kin}$ ($30-400$~K) with 100 logarithmic intervals each. The best-fit model gives the value of the $T_{\rm kin} = 308_{218}^{400}$~K, $n_{\rm H_2} = 1_{0.03}^{100}\times 10^{11}$~cm$^{-3}$ and $ N_{\rm HDO}=6.8_{4.6}^{6.8}\times10^{13}$~cm$^{-2}$, where we also provide the values corresponding to the $\pm 1 \sigma$ significance interval. Optical depths of the HDO transitions are $<1$. 
 
\begin{figure*}
    \centering
    \includegraphics[width=0.99\linewidth]{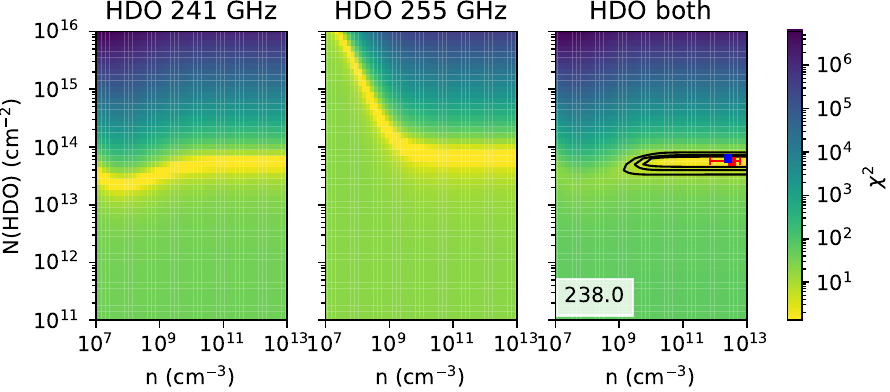}
    \caption{Results of non-LTE analysis for the HDO lines. The minimum of the $\chi^2=0.2$ value is shown by a blue dot. The average minimum $N_{\rm HDO}$ and $n_{\rm H_2}$ values with their standard deviations found by the bootstrapping procedure are shown by a red symbol with error bars.}
    \label{fig:HDO_RADEX}
\end{figure*}

Since the range of possible $T_{\rm kin}$ covers two hundred Kelvin, we wish to select some value and fix it to explore the degeneration of the parameters in 2D space of $n_{\rm H_2}$ and $N_{\rm HDO}$ values. From the analysis of more than forty lines of methanol in the spectrum of RCW\,120~S2 (Farafontova et al., in prep.), who found $T_{\rm kin}=238$~K for the hot core, we use this fixed value of $T_{\rm kin}$ below and perform the $\chi^2$-test in 2D space. Results of the fit and $\pm \sigma$ significance intervals, found with the $\chi^2$-test are shown in Fig.~\ref{fig:HDO_RADEX}. To assess how the $\chi^2$ minimum depends on the HDO line noise levels, we perform the bootstrapping procedure. Namely, we systematically varied the integrated intensities within their uncertainties given in Table~\ref{tab:gauss_sec}. By repeating this procedure 100 times, we determined the average minimum of the $\chi^2$ value and corresponding average parameters with their standard deviations. The average minimum $N_{\rm HDO}$ and $n_{\rm H_2}$ values with their standard deviations are also shown in Fig.~\ref{fig:HDO_RADEX} by red symbols with error bars.

Our non-LTE modeling reveals that while the $n_{\rm H_2}$ cannot be constrained, only the low limit was found, the $N_{\rm HDO}$ value is determined to within less than an order of magnitude. Analysis of the error bars obtained with the  bootstrapping procedure demonstrates that the best-fit model solution is reasonable. The value of the HDO column density is confined in a relatively thin interval $5.0\times10^{13} \leq N_{\rm HDO} \leq 6.3\times10^{13}$~cm$^{-2}$ with the best-fit value $N_{\rm HDO} = 6.3\times10^{13}$~cm$^{-2}$. The $\pm 3\sigma$ range for the HDO column density is $(3.9-7.9)\times10^{13}$~cm$^{-2}$. Dividing the $N_{\rm HDO}$ value by hydrogen column density, we obtain the relative abundance of HDO molecules $x_{\rm HDO} = 1.7\times10^{-9}$ in this hot gas.

\section{Discussion}

We detected two HDO lines and found intervals of appropriate physical parameters to excite these transitions. We note that the obtained range of H$_2$ number densities contains values much higher than critical densities of these transitions as \citet{2012MNRAS.420..699F} found typical value of the HDO critical density of $\sim 10^7$~cm$^{-3}$. Therefore, our initial LTE approach used to estimate the gas temperature is reasonable.

\begin{figure}
    \centering
    \includegraphics[width=0.7\linewidth]{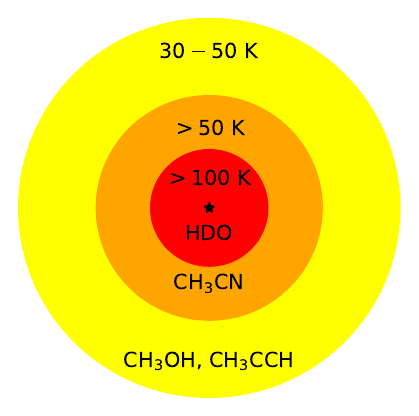}
    \caption{Schematic of the RCW\,120~S2 hot core (not to scale). There are at least three layers with different temperatures (indicated above the star symbol), where particular molecules are excited within according to this study and \citet{Kirsanova_2021, 2024paper1}. The dense warm envelope is shown in yellow and orange, but the hot core is shown in red colour.}
    \label{fig:plot_hot_hore}
\end{figure}

Combining our results with \citet{Kirsanova_2021} and \citet{2024paper1}, we plot a schematic of the RCW\,120~S2 in Fig.~\ref{fig:plot_hot_hore}. In this section, we discuss the chemical structure of the hot core. The $2_{\rm 1, 1}-2_{\rm 1, 2}$ and $5_{\rm 2, 3}-4_{\rm 3, 2}$ lines are approximately twice as large as those of the methanol, CH$_3$CCH, and CH$_3$CN lines originating in the dense warm envelope of RCW\,120~S2 \citet{2024paper1}. The HDO line width is consistent with those of the high-excitation CH$_3$CN lines ($12_K-11_K$ series with $K\geq 4$, $E_{\rm u} > 100$~K) originating in the hot core~\citep{Kirsanova_2021}. The agreement of the line widths supports our conclusion about the generation of the HDO emission in the hot core gas with $T_{\rm kin}\geq 100$~K.

Analysis of the methanol emission lines by \citet{2024paper1} showed the envelope is dense with $5.6\times10^5 \leq n_{\rm H_2} \leq 1.2\times10^6$~cm$^{-3}$ and warm $30 \leq T_{\rm kin} \leq 50$. Radial velocity of the component~3 agrees with the velocities of lines originating in the dense warm envelope. We suggest that the HDO spectra show no absorption features originating in the envelope, as the HDO lines have low optical depths. The absence of component~3 in the $5_{\rm 2, 3}-4_{\rm 3, 2}$ line profile reflects the insufficient thermal energy in the dense envelope to populate this high-excitation transition.

The para-H$_2^{18}$O $3_{\rm 1, 3}-2_{\rm 2, 0}$ line was not detected; consequently, only an upper limit on the water abundance could be established. With the $2\sigma$ level for the line intensity, we estimate $N_{\rm para-H_2^{18}O} \leq 6.6\times 10^{13}$~cm$^{-2}$. With the ortho-to-para ratio ${\rm o/p}=3$, we obtain $N_{\rm H_2^{18}O} \leq 2.6\times 10^{14}$~cm$^{-2}$. Taking into account the oxygen isotopic ratio for the galactocentric distance of RCW\,120 $^{16}{\rm O}/^{18}{\rm O}=460$\footnote{obtained from the fit $^{16}$O/$^{18}$O~$=(58.8\pm11.8)D_{\rm GC} + (37.1\pm82.6)$, where $D_{\rm GC}$ is the galactocentric distance, see \citep{Wilson_99}.}, we obtain $N_{\rm H_2O} \leq 1.2\times 10^{17}$~cm$^{-2}$ and $x_{\rm H_2O} \leq 3.2\times10^{-6}$. Therefore, the water deuterium fractionation in RCW\,120~S2 is $ {\rm HDO/H_2O} \geq 5.3\times10^{-4}$.

The upper limit on the water abundance in RCW\,120~S2 is one of the lowest values measured in other hot cores (see Fig.~21 in~\cite{vanDishoeck21}). Our value is only twice higher than the minimum detected water abundance in hot cores ($=1.7\times 10^{-6}$ in IRAS\,16272, see \cite{2016A&A...587A.139H}). All the cited abundance values were derived from models employing step-function profiles, where the inner hot regions near protostars exhibited abundances several orders of magnitude higher than the outer envelopes. The transition boundary corresponded to the 100~K radius surrounding YSOs. Considering the component~3 originating in the warm gas of $30 \leq T_{\rm kin} \leq 50$~K,  we also obtain a relative abundance of water lower than $10^{-6}$ by at least the order of magnitude. We will perform detailed chemical modeling after completing the identification of all detected spectral lines (Farafontova et al., in prep.).

If most oxygen were locked in water molecules that were predominantly frozen onto dust grains, the resulting gas-phase water abundance in hot cores would be $\approx 10^{-4}$ relative to hydrogen. Observational studies of high-mass sources including hot cores typically find $5\times 10^{-6} \leq x_{\rm H_2O} \leq 10^{-4}$ \citep{vanDishoeck21} with a few exceptions: W43-MM1 \citep{2016A&A...587A.139H} and W3\,IRS5 \citep{2010A&A...521L..37C} where relative water abundance is $\approx 10^{-4}$. Thus, the {\it Herschel} observations revealed a problem of 'dry' hot cores. RCW\,120~S2 represents one of the most water-depleted hot cores ever observed, and may in fact be the driest known example to date. Non-detection of the H$_2^{18}$O line at 203~GHz supports our conclusions.

\citet{2022ApJ...933...64S} found the HDO line luminosity is largely dependent on the source luminosity $L_{\rm bol}$ determined by temperature. Using the HDO line at 241~GHz and assuming the hot core as a point source, we derived $L_{\rm HDO}=8.5\times10^{-3}$~$L_\odot$ and found that RCW\,120~S2 has one of the lowest $L_{\rm HDO}$ and $L_{\rm bol}$ compared to other known hot cores. On a graph plotting these two values in the cited study, it appears in the bottom-left corner. We conclude the hot core has low $x_{\rm HDO}$ because the central heat source is not strong enough to fully evaporate the icy mantles on dust grains.

We found in Sec.~\ref{sec:results} that the $2_{\rm 1, 1}-2_{\rm 1, 2}$ and $5_{\rm 2, 3}-4_{\rm 3, 2}$ lines are optically thin. Their double-peaked shapes, where both peaks are about equally strong, might be caused by either gas infalling toward the central source  or possibly outflowing gas \citep[see an example e.~g. in][]{UFN}. The best-fit RADEX solution gives $\tau < 1$ for both the HDO lines. Without clear detections of either inverse P-Cygni (infall) or regular P-Cygni (outflow) signatures in the line profiles, those would be natural for the optically thick lines, we cannot establish the gas dynamics in this hot core with confidence. Signatures for both types of kinematics were found in RCW\,120~S2. \citet{2023MNRAS.520..751K} detected a skewed profile in the $^{13}$CO(6--5) line, consistent with gas infall kinematics. In contrast, \citet{2024paper1} identified a molecular outflow in this region through methanol transitions with lower upper-state energies ($E_{\rm u}$) than those of the HDO lines we observed. Both types of gas kinematics were found in various hot cores previously by e.~g.~\citet{2016A&A...587A.139H}, who analysed plenty of water emission lines. Given the absence of additional HDO or water transitions in our dataset, we defer the investigation of gas kinematics to future studies.

\section{Conclusions}

We detected two HDO lines at 241 and 255~GHz toward young hot core RCW\,120~S2. These lines are originating in a gas with  temperature more than two hundred of Kelvin and gas number density $\geq 10^9$~cm$^{-3}$ representing a hot core. The core is surrounded by dense warm envelope with $T_{\rm kin} =30-50$~K and gas number density $n_{\rm H_2} \approx 10^6$~cm$^{-3}$. The HDO column density in the hot core ranges from $(3.9-7.9)\times10^{13}$~cm$^{-3}$ with a most probable value $6.3\times10^{13}$~cm$^{-3}$. The HDO abundance relative to hydrogen nuclei is $1.7\times10^{-9}$.     

Non-detection of H$_2^{18}$O line allows estimating the upper limit of water column density $N_{\rm H_2O}\leq 1.2\times10^{17}$~cm$^{-2}$ and the water deuterium fractionation as $N_{\rm HDO}/N_{\rm H_2^{18}O} \geq 5.3\times10^{-4}$.  RCW\,120~S2 represents one of the most water-depleted hot cores ever observed, and may in fact be the driest known example to date. In the RCW\,120~S2 hot core, the central heat source is not strong enough to fully evaporate the icy mantles on dust grains.

\section*{Acknowledgments}

We are thankful to Ya. N. Pavlyuchenkov for a useful discussion. We are also thankful to the anonymous referee, whose comments and suggestions helped us to improve the study.

\bibliographystyle{mnras}
\bibliography{HDO_APEX} 

\begin{thebibliography}{}
\makeatletter
\relax
\def\mn@urlcharsother{\let\do\@makeother \do\$\do\&\do\#\do\^\do\_\do\%\do\~}
\def\mn@doi{\begingroup\mn@urlcharsother \@ifnextchar [ {\mn@doi@}
  {\mn@doi@[]}}
\def\mn@doi@[#1]#2{\def\@tempa{#1}\ifx\@tempa\@empty \href
  {http://dx.doi.org/#2} {doi:#2}\else \href {http://dx.doi.org/#2} {#1}\fi
  \endgroup}
\def\mn@eprint#1#2{\mn@eprint@#1:#2::\@nil}
\def\mn@eprint@arXiv#1{\href {http://arxiv.org/abs/#1} {{\tt arXiv:#1}}}
\def\mn@eprint@dblp#1{\href {http://dblp.uni-trier.de/rec/bibtex/#1.xml}
  {dblp:#1}}
\def\mn@eprint@#1:#2:#3:#4\@nil{\def\@tempa {#1}\def\@tempb {#2}\def\@tempc
  {#3}\ifx \@tempc \@empty \let \@tempc \@tempb \let \@tempb \@tempa \fi \ifx
  \@tempb \@empty \def\@tempb {arXiv}\fi \@ifundefined
  {mn@eprint@\@tempb}{\@tempb:\@tempc}{\expandafter \expandafter \csname
  mn@eprint@\@tempb\endcsname \expandafter{\@tempc}}}

\bibitem[\protect\citeauthoryear{{Belitsky} et~al.,}{{Belitsky}
  et~al.}{2018}]{2018A&A...611A..98B}
{Belitsky} V.,  et~al., 2018, \mn@doi [\aap] {10.1051/0004-6361/201731883},
  \href {https://ui.adsabs.harvard.edu/abs/2018A&A...611A..98B} {611, A98}

\bibitem[\protect\citeauthoryear{{Chavarr{\'\i}a} et~al.,}{{Chavarr{\'\i}a}
  et~al.}{2010}]{2010A&A...521L..37C}
{Chavarr{\'\i}a} L.,  et~al., 2010, \mn@doi [\aap]
  {10.1051/0004-6361/201015113}, \href
  {https://ui.adsabs.harvard.edu/abs/2010A&A...521L..37C} {521, L37}

\bibitem[\protect\citeauthoryear{{Cuppen} \& {Herbst}}{{Cuppen} \&
  {Herbst}}{2007}]{2007ApJ...668..294C}
{Cuppen} H.~M.,  {Herbst} E.,  2007, \mn@doi [\apj] {10.1086/521014}, \href
  {https://ui.adsabs.harvard.edu/abs/2007ApJ...668..294C} {668, 294}

\bibitem[\protect\citeauthoryear{Erb}{Erb}{2024}]{erb_2024}
Erb D.,  2024, pybaselines: A Python library of algorithms for the baseline
  correction of experimental data, \mn@doi{10.5281/zenodo.10676584}, \url
  {https://doi.org/10.5281/zenodo.10676584}

\bibitem[\protect\citeauthoryear{{Faure}, {Wiesenfeld}, {Scribano}  \&
  {Ceccarelli}}{{Faure} et~al.}{2012}]{2012MNRAS.420..699F}
{Faure} A.,  {Wiesenfeld} L.,  {Scribano} Y.,   {Ceccarelli} C.,  2012, \mn@doi
  [\mnras] {10.1111/j.1365-2966.2011.20081.x}, \href
  {https://ui.adsabs.harvard.edu/abs/2012MNRAS.420..699F} {420, 699}

\bibitem[\protect\citeauthoryear{{Figueira} et~al.,}{{Figueira}
  et~al.}{2017}]{2017A&A...600A..93F}
{Figueira} M.,  et~al., 2017, \mn@doi [\aap] {10.1051/0004-6361/201629379},
  \href {https://ui.adsabs.harvard.edu/abs/2017A&A...600A..93F} {600, A93}

\bibitem[\protect\citeauthoryear{{Figueira}, {Bronfman}, {Zavagno}, {Louvet},
  {Lo}, {Finger}  \& {Rod{\'o}n}}{{Figueira}
  et~al.}{2018}]{2018A&A...616L..10F}
{Figueira} M.,  {Bronfman} L.,  {Zavagno} A.,  {Louvet} F.,  {Lo} N.,  {Finger}
  R.,   {Rod{\'o}n} J.,  2018, \mn@doi [\aap] {10.1051/0004-6361/201832930},
  \href {https://ui.adsabs.harvard.edu/abs/2018A&A...616L..10F} {616, L10}

\bibitem[\protect\citeauthoryear{{Goldsmith} \& {Langer}}{{Goldsmith} \&
  {Langer}}{1999}]{Goldsmith1999}
{Goldsmith} P.~F.,  {Langer} W.~D.,  1999, \mn@doi [\apj] {10.1086/307195},
  \href {https://ui.adsabs.harvard.edu/abs/1999ApJ...517..209G} {517, 209}

\bibitem[\protect\citeauthoryear{{G{\"u}sten}, {Nyman}, {Schilke}, {Menten},
  {Cesarsky}  \& {Booth}}{{G{\"u}sten} et~al.}{2006}]{2006A&A...454L..13G}
{G{\"u}sten} R.,  {Nyman} L.~{\r{A}}.,  {Schilke} P.,  {Menten} K.,  {Cesarsky}
  C.,   {Booth} R.,  2006, \mn@doi [\aap] {10.1051/0004-6361:20065420}, \href
  {https://ui.adsabs.harvard.edu/abs/2006A&A...454L..13G} {454, L13}

\bibitem[\protect\citeauthoryear{{Herpin} et~al.,}{{Herpin}
  et~al.}{2016}]{2016A&A...587A.139H}
{Herpin} F.,  et~al., 2016, \mn@doi [\aap] {10.1051/0004-6361/201527786}, \href
  {https://ui.adsabs.harvard.edu/abs/2016A&A...587A.139H} {587, A139}

\bibitem[\protect\citeauthoryear{{Jacq}, {Walmsley}, {Henkel}, {Baudry},
  {Mauersberger}  \& {Jewell}}{{Jacq} et~al.}{1990}]{Jacq_1990}
{Jacq} T.,  {Walmsley} C.~M.,  {Henkel} C.,  {Baudry} A.,  {Mauersberger} R.,
  {Jewell} P.~R.,  1990, \aap, \href
  {https://ui.adsabs.harvard.edu/abs/1990A&A...228..447J} {228, 447}

\bibitem[\protect\citeauthoryear{{Kalenskii} \& {Kurtz}}{{Kalenskii} \&
  {Kurtz}}{2016}]{2016ARep...60..702K}
{Kalenskii} S.~V.,  {Kurtz} S.,  2016, \mn@doi [Astronomy Reports]
  {10.1134/S1063772916080047}, \href
  {https://ui.adsabs.harvard.edu/abs/2016ARep...60..702K} {60, 702}

\bibitem[\protect\citeauthoryear{{Kirsanova}, {Salii}, {Kalenskii}, {Wiebe},
  {Sobolev}  \& {Boley}}{{Kirsanova} et~al.}{2021}]{Kirsanova_2021}
{Kirsanova} M.~S.,  {Salii} S.~V.,  {Kalenskii} S.~V.,  {Wiebe} D.~S.,
  {Sobolev} A.~M.,   {Boley} P.~A.,  2021, \mn@doi [\mnras]
  {10.1093/mnras/stab499}, \href
  {https://ui.adsabs.harvard.edu/abs/2021MNRAS.503..633K} {503, 633}

\bibitem[\protect\citeauthoryear{{Kirsanova}, {Pavlyuchenkov}, {Olofsson},
  {Semenov}  \& {Punanova}}{{Kirsanova} et~al.}{2023}]{2023MNRAS.520..751K}
{Kirsanova} M.~S.,  {Pavlyuchenkov} Y.~N.,  {Olofsson} A.~O.~H.,  {Semenov}
  D.~A.,   {Punanova} A.~F.,  2023, \mn@doi [\mnras] {10.1093/mnras/stac3737},
  \href {https://ui.adsabs.harvard.edu/abs/2023MNRAS.520..751K} {520, 751}

\bibitem[\protect\citeauthoryear{{Kirsanova} et~al.,}{{Kirsanova}
  et~al.}{2025}]{UFN}
{Kirsanova} M.~S.,  et~al., 2025, \mn@doi [Phys. Usp.]
  {10.3367/UFNe.2024.08.039744}, 68

\bibitem[\protect\citeauthoryear{{Messer}, {De Lucia}  \& {Helminger}}{{Messer}
  et~al.}{1984}]{1984JMoSp.105..139M}
{Messer} J.~K.,  {De Lucia} F.~C.,   {Helminger} P.,  1984, \mn@doi [Journal of
  Molecular Spectroscopy] {10.1016/0022-2852(84)90109-7}, \href
  {https://ui.adsabs.harvard.edu/abs/1984JMoSp.105..139M} {105, 139}

\bibitem[\protect\citeauthoryear{{M{\"u}ller}, {Thorwirth}, {Roth}  \&
  {Winnewisser}}{{M{\"u}ller} et~al.}{2001}]{2001A&A...370L..49M}
{M{\"u}ller} H.~S.~P.,  {Thorwirth} S.,  {Roth} D.~A.,   {Winnewisser} G.,
  2001, \mn@doi [\aap] {10.1051/0004-6361:20010367}, \href
  {https://ui.adsabs.harvard.edu/abs/2001A&A...370L..49M} {370, L49}

\bibitem[\protect\citeauthoryear{{Oba}, {Watanabe}, {Hama}, {Kuwahata},
  {Hidaka}  \& {Kouchi}}{{Oba} et~al.}{2012}]{2012ApJ...749...67O}
{Oba} Y.,  {Watanabe} N.,  {Hama} T.,  {Kuwahata} K.,  {Hidaka} H.,   {Kouchi}
  A.,  2012, \mn@doi [\apj] {10.1088/0004-637X/749/1/67}, \href
  {https://ui.adsabs.harvard.edu/abs/2012ApJ...749...67O} {749, 67}

\bibitem[\protect\citeauthoryear{{Pickett}, {Poynter}, {Cohen}, {Delitsky},
  {Pearson}  \& {M{\"u}ller}}{{Pickett} et~al.}{1998}]{1998JQSRT..60..883P}
{Pickett} H.~M.,  {Poynter} R.~L.,  {Cohen} E.~A.,  {Delitsky} M.~L.,
  {Pearson} J.~C.,   {M{\"u}ller} H.~S.~P.,  1998, \mn@doi [\jqsrt]
  {10.1016/S0022-4073(98)00091-0}, \href
  {https://ui.adsabs.harvard.edu/abs/1998JQSRT..60..883P} {60, 883}

\bibitem[\protect\citeauthoryear{{Plakitina}, {Kirsanova}, {Kalenskii}, {Salii}
   \& {Wiebe}}{{Plakitina} et~al.}{2024}]{2024paper1}
{Plakitina} K.~V.,  {Kirsanova} M.~S.,  {Kalenskii} S.~V.,  {Salii} S.~V.,
  {Wiebe} D.~S.,  2024, Astrophysical Bulletin, 79, 235

\bibitem[\protect\citeauthoryear{{Russeil}}{{Russeil}}{2003}]{2003A&A...397..133R}
{Russeil} D.,  2003, \mn@doi [\aap] {10.1051/0004-6361:20021504}, \href
  {https://ui.adsabs.harvard.edu/abs/2003A&A...397..133R} {397, 133}

\bibitem[\protect\citeauthoryear{{Sewi{\l}o} et~al.,}{{Sewi{\l}o}
  et~al.}{2022}]{2022ApJ...933...64S}
{Sewi{\l}o} M.,  et~al., 2022, \mn@doi [\apj] {10.3847/1538-4357/ac6de1}, \href
  {https://ui.adsabs.harvard.edu/abs/2022ApJ...933...64S} {933, 64}

\bibitem[\protect\citeauthoryear{{Tielens} \& {Hagen}}{{Tielens} \&
  {Hagen}}{1982}]{1982A&A...114..245T}
{Tielens} A.~G.~G.~M.,  {Hagen} W.,  1982, \aap, \href
  {https://ui.adsabs.harvard.edu/abs/1982A&A...114..245T} {114, 245}

\bibitem[\protect\citeauthoryear{{Van der Tak}, {Black}, {Sch{\"o}ier},
  {Jansen}  \& {van Dishoeck}}{{Van der Tak} et~al.}{2007}]{vanderTak_2007}
{Van der Tak} F.~F.~S.,  {Black} J.~H.,  {Sch{\"o}ier} F.~L.,  {Jansen} D.~J.,
   {van Dishoeck} E.~F.,  2007, \mn@doi [\aap] {10.1051/0004-6361:20066820},
  \href {https://ui.adsabs.harvard.edu/abs/2007A&A...468..627V} {468, 627}

\bibitem[\protect\citeauthoryear{Wall \& Jenkins}{Wall \&
  Jenkins}{2003}]{wall_jenkins_2003}
Wall J.~V.,  Jenkins C.~R.,  2003, Practical Statistics for Astronomers.
Cambridge Observing Handbooks for Research Astronomers, Cambridge University
  Press, \mn@doi{10.1017/CBO9780511536618}

\bibitem[\protect\citeauthoryear{{Wilson}}{{Wilson}}{1999}]{Wilson_99}
{Wilson} T.~L.,  1999, \mn@doi [Reports on Progress in Physics]
  {10.1088/0034-4885/62/2/002}, \href
  {http://adsabs.harvard.edu/abs/1999RPPh...62..143W} {62, 143}

\bibitem[\protect\citeauthoryear{{de Lucia}, {Helminger}, {Cook}  \&
  {Gordy}}{{de Lucia} et~al.}{1972}]{1972PhRvA...6.1324D}
{de Lucia} F.~C.,  {Helminger} P.,  {Cook} R.~L.,   {Gordy} W.,  1972, \mn@doi
  [\pra] {10.1103/PhysRevA.6.1324}, \href
  {https://ui.adsabs.harvard.edu/abs/1972PhRvA...6.1324D} {6, 1324}

\bibitem[\protect\citeauthoryear{{van Dishoeck}, {Herbst}  \& {Neufeld}}{{van
  Dishoeck} et~al.}{2013}]{2013ChRv..113.9043V}
{van Dishoeck} E.~F.,  {Herbst} E.,   {Neufeld} D.~A.,  2013, \mn@doi [Chemical
  Reviews] {10.1021/cr4003177}, \href
  {https://ui.adsabs.harvard.edu/abs/2013ChRv..113.9043V} {113, 9043}

\bibitem[\protect\citeauthoryear{{van Dishoeck} et~al.,}{{van Dishoeck}
  et~al.}{2021}]{vanDishoeck21}
{van Dishoeck} E.~F.,  et~al., 2021, \mn@doi [\aap]
  {10.1051/0004-6361/202039084}, \href
  {https://ui.adsabs.harvard.edu/abs/2021A&A...648A..24V} {648, A24}

\makeatother
\end{thebibliography}

\bsp	
\label{lastpage}
\end{document}